\documentclass[sigconf,nonacm]{acmart}
\AtBeginDocument{%
  \providecommand\BibTeX{{%
    \normalfont B\kern-0.5em{\scshape i\kern-0.25em b}\kern-0.8em\TeX}}}

\usepackage{multirow}
\usepackage{balance}

\begin{document}

\title{You too Brutus! Trapping Hateful Users in Social Media:\\ Challenges, Solutions \& Insights}

\author{Mithun Das, Punyajoy Saha, Ritam Dutt, Pawan Goyal, Animesh Mukherjee, Binny Mathew}
\affiliation{%
  \institution{Department of Computer Science \& Engineering,\\  Indian Institute of Technology, Kharagpur}
  \state{West Bengal}
  \country{India -- 721302}
}
\email{{mithundas,punyajoys,binnymathew}@iitkgp.ac.in,{animeshm,pawang}@cse.iitkgp.ac.in}

\newcommand{\red}[1]{\textcolor{red}{#1}}
\renewcommand{\shortauthors}{Das, et al.}

\begin{abstract}
Hate speech is regarded as one of the crucial issues plaguing the online social media. The current literature on hate speech detection leverages primarily the textual content to find hateful posts and subsequently identify hateful users. However, this methodology disregards the social connections between users. In this paper, we run a detailed exploration of the problem space and investigate an array of models ranging from purely textual to graph based to finally semi-supervised techniques using Graph Neural Networks (GNN) that utilize both textual and graph-based features. We run exhaustive experiments on two datasets -- \textbf{Gab}\footnote{\label{gab_site}\url{gab.com}}, which is loosely moderated and \textbf{Twitter}\footnote{\label{twitter_site}\url{twitter.com}}, which is strictly moderated. Overall the AGNN model achieves 0.791 macro F1-score on the Gab dataset and 0.780 macro F1-score on the Twitter dataset using only 5\% of the labeled instances, considerably outperforming all the other models including the fully supervised ones. We perform detailed error analysis on the best performing text and graph based models and observe that hateful users \textit{have unique network neighborhood signatures} and the AGNN model benefits by paying attention to these signatures. This property, as we observe, also allows the model to generalize well across platforms in a zero-shot setting. Lastly, we utilize the best performing GNN model to analyze the evolution of hateful users and their targets over time in Gab.
\end{abstract}




\maketitle

\section{Introduction}
Hate speech is regarded as one of the major issues plaguing the online social media, necessitating its detection as a crucial task. Hate speech detection itself is a challenging task with literature including techniques such as dictionary-based~\cite{guermazi2007using}, distributional semantics~\cite{djuric2015hate}, multi-feature~\cite{salminen2018anatomy} and neural networks~\cite{Badjatiya:2017:DLH:3041021.3054223}.

The majority of these systems rely solely on or mostly on the textual content of the post. However, hate speech is highly subjective, reliant on temporal, social and historical context, and occurs sparsely~\cite{schmidt2017survey}. Hate speech is often crafted in a  subtle manner and cannot be precisely identified solely using text-based features. Furthermore, hate speech detection from text is a closed-loop problem~\cite{macavaney2019hate}. Since the hateful users are conscious of the  detection systems, they will try to include deliberate spelling errors or code names of the targets to spread hate speech. Hence, a better way will be to consider the users' profile, interests, posting behaviour, and people they are linked to in the online social networks (OSNs). Recently, there have been efforts to leverage these user information to increase the performance~\cite{qian2018leveraging} but a critical analysis of such network based methods is lacking in the literature. In this work, we explore several graph based methods to detect hateful users in online social networks while focusing on the challenges present while doing so. We further provide insights and observations based on the analysis and provide some suggestions about the future research directions.

\noindent\textbf{Challenges}: While the motivation to escalate the detection problem from the post level to the user level looks apparent, one has to note that it is an extremely tedious task to get large scale annotated data at the user level. This is since the annotators now would need to look through all/a majority of the posts of a user to designate a user as hateful. Therefore the option of adapting existing popular supervised text-based models to detect hateful users is ruled out since these are data hungry and would typically not perform well in low data settings.

\noindent\textbf{Solution}: Recently an array of research have tried to use graph based algorithms; however, most of these do not perform an extensive comparison of supervised algorithms~\cite{mishra-etal-2018-author}. In this paper we attempt to bridge this gap by presenting a rigorous evaluation of the graph algorithms, through few shot and cross platform test scenarios. We utilize several text and graph based methods for identifying hateful users. These include LSTM, doc2vec, BERT, node2vec, DeepWalk, GCN, GraphSAGE etc. We test these models on two different datasets (Twitter and Gab). To test the model's capability to work with lesser data, we test the model's performance on 5\%, 10\%, 15\%, 20\%, 50\%, and 80\% of the training data. In Gab dataset, users are nodes and the edges represent the follower-followee relationship. In Twitter, users are nodes and the edges represent the retweet information. We thus experiment on two types of network and report the performance.

Our \textbf{key contributions} are noted below.
\begin{itemize}

\item We create a dataset of 423 hateful users and 375 non-hateful users from the social media platform Gab where each user was adjudged to be hateful or not based on the profile information. For Twitter, we make use of the data already made available by Ribeiro et al.~\shortcite{ribeiro2018characterizing}. We have made the Gab dataset along with all the codes public for advancing research in hate speech\footnote{\url{https://github.com/hate-alert/Hateful-users-detection}}.

\item We explore several supervised, unsupervised and semi supervised machine learning models, including the state-of-the-art deep learning models, to classify users as hateful and non-hateful.
\item We perform detailed error analysis on the best performing text based and graph based model.
\item We apply the trained model to label the entire Gab dataset and perform a post-facto analysis of the evolution of hatred across certain target communities.
\end{itemize}

The \textbf{most important observations} that we make are as follows.
\begin{itemize}

    \item We observe that semi-supervised approaches using GNN that leverage both textual features and social connections between users significantly outperform other models. For the Gab dataset, the best GNN model achieves a macro F1-Score of 0.791 and for the Twitter dataset, it achieves a macro F1-Score of 0.780 with \textbf{only 5\% of the labelled instances}. The performance is comparable with other supervised machine learning classifiers like LSTM and doc2vec, that use the entire set of labelled instances for training.
    \item On cross-platform zero shot evaluation, we observe that AGNN performs better as compared to the best text-based models. The attention on the neighborhood structure (learnt from one dataset) seems to capture the signature characteristic of the users and constitutes as one of the most effective ingredient for predictions given an unseen dataset.
    \item A detailed error analysis reveals that when words from the hate lexicon are infrequent in a user's post, standard supervised models (e.g., doc2vec+LR) are unable to detect if the user is hateful; in contrast, graph based semi-supervised models (e.g., AGNN) are able to detect such a user correctly by making use of the hateful influential nodes (read users) in the neighbourhood of that user.
    \item Post-facto analysis of the machine-labelled Gab data shows that ethnic and religious groups like Blacks, Jews and Muslims face an ever increasing hatred from the community; i.e., as the site is getting older, so is the hatred blowing up against these communities. 
\end{itemize} 

\section{Related work}

\noindent\textbf{Hate speech detection}: Hate speech is a complex phenomenon, intrinsically associated to relationships among groups, and also relying on linguistic nuances~\cite{fortuna2018survey}. The public expression of such hate speech has been shown to devalue members of the minority community~\cite{greenberg1985effect}. To tackle this issue, researchers have developed methods to detect such hateful content. Some of the initial works of detection of hate speech relied on lexicons \cite{warner2012detecting,gitari2015lexicon}. Liu and Forss \shortcite{liu2015new} incorporated LDA topic modelling for improving the performance of the hate speech detection task. Saleem et al.~\shortcite{saleem2017web} proposed an approach to detect hateful speech using self-identifying hate communities as training data for hate speech classifiers thus bypassing the expensive annotation process. 

\noindent\textbf{Deep learning approaches}: Recently, larger datasets for hate speech detection have been made available~\cite{davidson2017automated,founta2018large,de2018hate}. Most of these datasets have hate speech class as the minority. Researchers have also started using deep learning methods \cite{zhang2018detecting} and graph embedding techniques~\cite{ribeiro2018characterizing} to detect hate speech. Badjatiya et al.~\shortcite{Badjatiya:2017:DLH:3041021.3054223} applied several deep learning architectures and improved the benchmark score by $\sim$18 F1 points. Zhang et al.~\shortcite{zhang2018detecting} used novel deep neural models to improve the results on 6 out of 7 datasets.

\noindent\textbf{Hate speech at user level}: While most of the computational approaches focus on detecting if a given text contains hate speech, very few works focus on detecting this at the user level. Detecting hate speech at user level will allow algorithms to use additional dimensions such as user activity and connection which could help in improving the classifier performance~\cite{ribeiro2018characterizing}. Qian et al.~\shortcite{qian2018leveraging} proposes a model that learns intra-user and inter-user representations for hate speech detection.

In our paper, we explore an array of methods to detect hateful accounts on Gab and Twitter. The work by Ribeiro et al.~\shortcite{ribeiro2018characterizing} is closest to ours and hence we describe it briefly here. Ribeiro et al.~\shortcite{ribeiro2018characterizing} builds a Twitter retweet graph and uses a graph embedding approach to detect hateful users. They collect and annotate around 5K users on Twitter and characterize their Twitter accounts. The authors then employ a node embedding algorithm (GraphSAGE~\cite{GraphSAGE}),  which  exploits  the  graph  structure and show that it outperforms content-based approaches for hateful user detection. Our work, on the other hand, explores the performance of supervised/semi-supervised models that use both the content and the network structure to detect hateful users in Gab and Twitter. The uniqueness of our models lies in the use of very small number of labelled instances. Further we add two more novelties - (i) an extensive error analysis which shows how the neighborhood characteristic of hateful users benefits GNN models and (ii) predictions in the zero-shot setting.

\section{Methodology}
In this section, we present a suite of models for hateful user detection. These range from text based approaches to graph based approaches to finally text+graph based approaches. The methods are enumerated below. This pipeline of efforts portrays a \textit{hitherto unreported complete picture} of how the different online attributes can be effective in determining hateful users.

\begin{figure}[h]
  \centering
	\includegraphics[width=\linewidth]{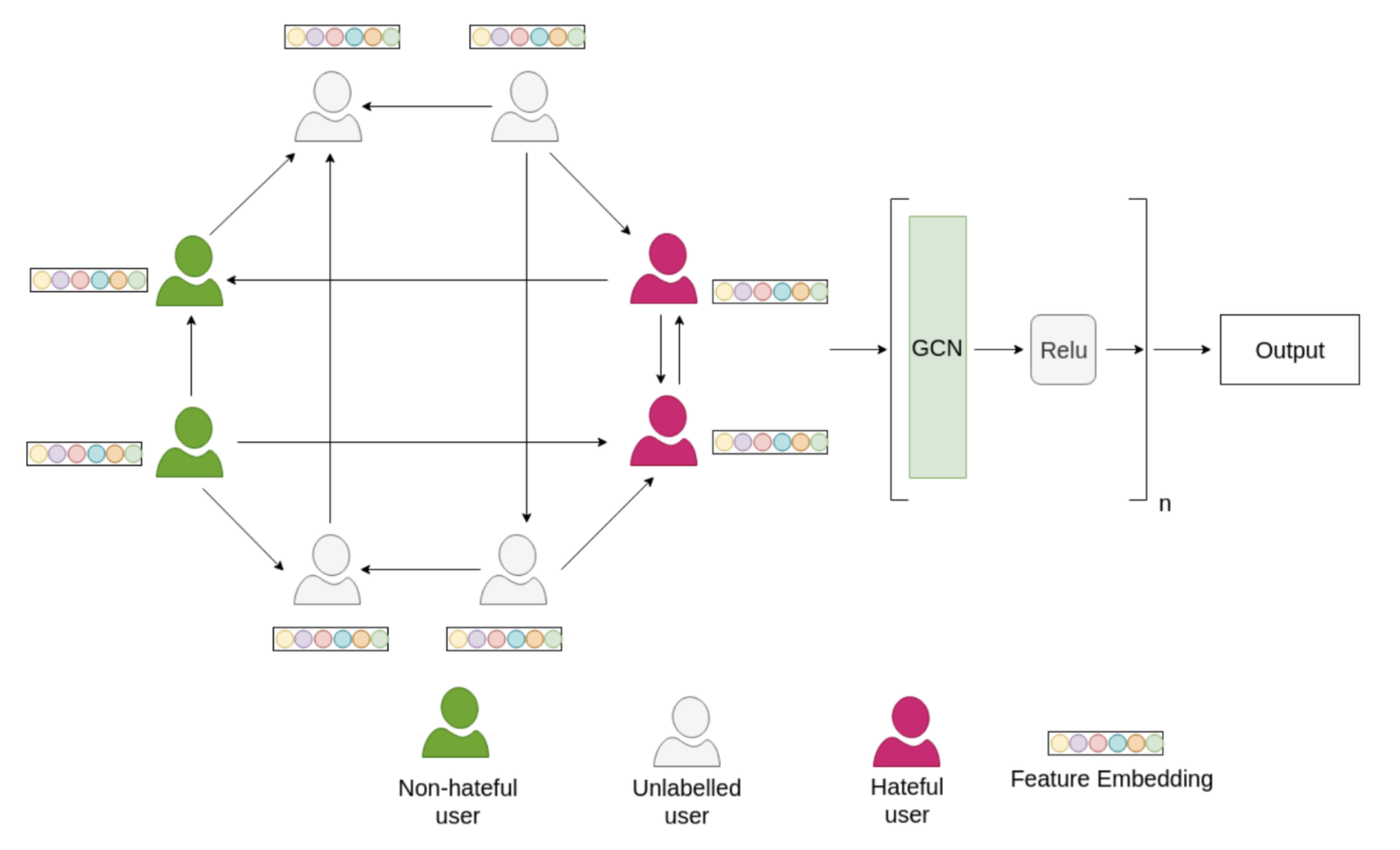}
    \caption{Schematic representation of a GNN classifier (in this case GCN) which takes in as input the social network of the users and the corresponding feature embeddings of each user. After applying ``n'' GCN filters on the input data, we obtain the output i.e, the  probability of the user being hateful or not. }
	\label{fig:text-based-classification}
\end{figure}

\begin{figure}[h]
  \centering
  \includegraphics[width=\linewidth]{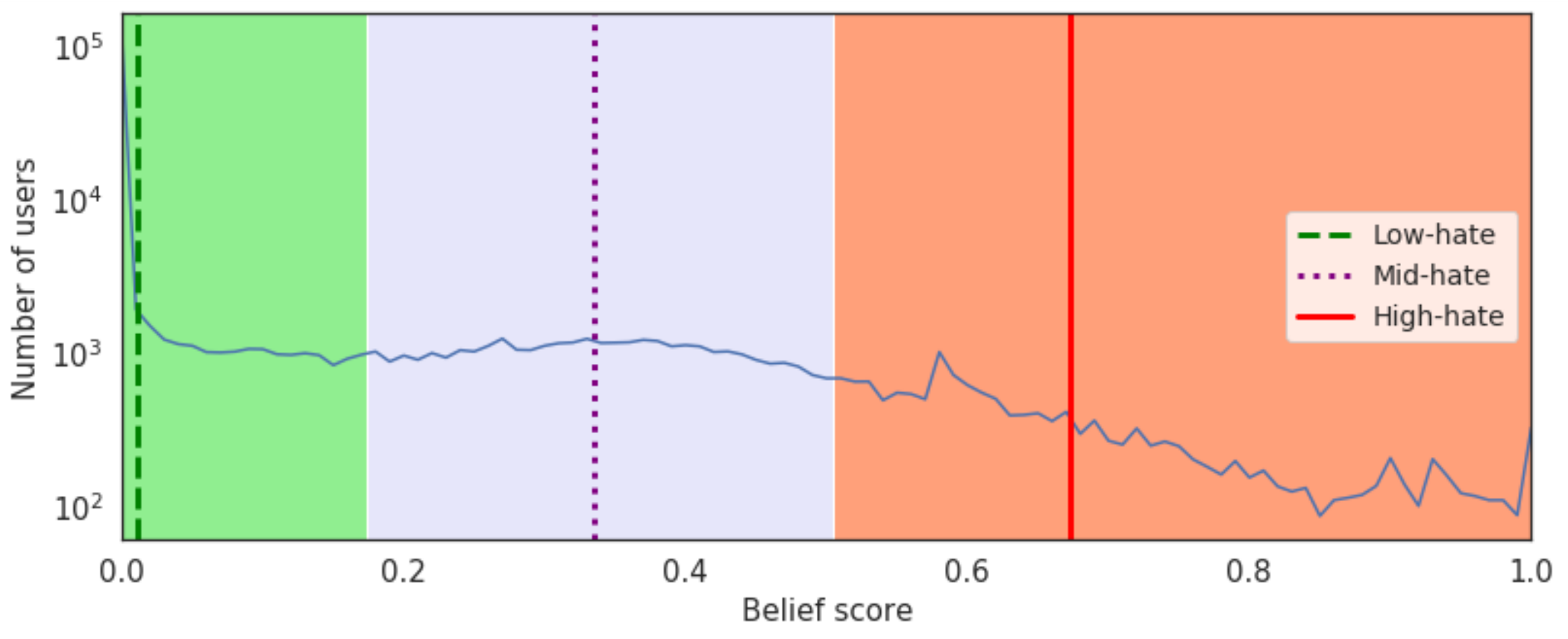}
    \caption{We show the distribution of the belief scores of the users. The dashed lines in the diagram correspond to the centroids of ``low'', ``medium'' and ``high'' hate intensity.}
  \label{fig:belief-score-dist}
\end{figure}

\subsection{Text-based classification}
Here the idea is to classify users as hateful and non-hateful solely based on the content of their posts. We apply standard pre-processing techniques on the post to remove urls, mentions, hashtags, emoticons and other stray characters. 
Finally, we concatenate all the posts of an user into a single \textit{paragraph}/\textit{document} and provide it as an input to the classifiers. We experiment with the following text-classifiers. 

\noindent\textbf{fastText:} We leverage the pre-trained fastText embedding by Grave et al.~\cite{grave2018learning} which has been trained on Common Crawl and Wikipedia. Each user's document has been represented as a 100 dimensional vector. We use logistic regression as the classifier.

\noindent\textbf{Glove}: We use the  pre-trained Glove embeddings~\cite{pennington2014glove} which has been trained on 2B tweets, to represent each word as a 100 dimensional vector. We then represent a user as the mean of all the Glove embeddings of all the words in his/her posts. We use logistic regression as the classifier.  

\noindent\textbf{LSTM}: We experiment with standard LSTM~\cite{lstm} and neural models with random embeddings, applied over the user document.
The loss is set to binary cross-entropy. The model is run for 10 epochs with Adam optimizer using default parameters of keras\footnote{https://bit.ly/2zBs87Z}.

\noindent\textbf{Doc2vec}: We apply the doc2vec model~\cite{doc2vec-paper} on the user document to generate a default 100 dimensional document embedding for each user. We use logistic regression as the classifier. We use the default hyper-parameters of the doc2vec implementation available in gensim~\cite{rehurek2010software}.

\noindent\textbf{BERT}: In order to see if contextual embeddings can better represent a user, we fine-tuned the already pre-trained BERT model. For finetuning BERT we follow a setup similar to Doc-BERT \cite{adhikari2019docbert}. For each user in the labelled set, we combine all the post of that user and consider it as a document to be used in the training of the model. While tokenizing a document we consider the first 512 tokens which is the limit of the input that can be passed to BERT. We train the BERT model for 10 epochs with a default learning rate of 2e-5 and store the results at best validation score. 

\noindent\textbf{TSVM}: We use the Transductive Support Vector Machines (TSVM) model~\cite{TSVM-paper} for semi-supervised classification. Applying doc2vec \cite{doc2vec-paper} on the user document yields the user feature vector. TSVM aims to learn the manifold  between the hateful and non-hateful users by leveraging the feature vectors of both the labelled and unlabelled users. We use the default hyper-parameters of the implementation\footnote{https://bit.ly/3iviua2}.

\subsection{Network embeddings}
Research on hate speech in social media has revealed that hateful users are densely connected in the network~\cite{mathew2019hate}. They exhibit a strong degree of homophily and have high reciprocity values~\cite{ribeiro2018characterizing,hate-speech-websci-19}. Consequently the network structure might provide additional insights for detecting hateful users. Network or node embeddings enable us to project the nodes to a lower-dimensional latent embedding space, while preserving its network characteristics and have been instrumental in node classification~\cite{8392745}. We explore the embeddings generated by DeepWalk and Node2vec in this work.

\noindent\textbf{DeepWalk:} DeepWalk~\cite{deepwalk-paper} performs random walk on the network to generate a sequence of nodes. The sequence of nodes can be imagined to be a sentence with each node representing a word. Applying word2vec on these simulated sentences generates an embedding for each node. We apply DeepWalk~\cite{deepwalk-paper} on the network $G$ to learn a 128 dimensional embedding for each node. We apply logistic regression on the learned representation of the nodes to classify the users. The hyper-parameters are 10 random walks per node with a walk-length of 80, and a window-size of 10.

\noindent\textbf{Node2vec:} The Node2vec algorithm learns low-dimensional representation for nodes in a graph that maximizes the likelihood of preserving network neighborhoods of nodes. We apply Node2vec \cite{Node2vec-kdd2016} on the network $G$ to learn a 128 dimensional embedding for each node. We use the default hyper-parameters to generate the node embedding. We apply logistic regression as a classifier.

\subsection{Graph neural networks (GNN)} Node embeddings are typically shallow encoders; they do not allow parameter sharing nor do they incorporate node features~\cite{GraphSAGE}. To overcome these limitations, we have employed Graph Neural Networks (GNN). GNN operates on a graph and can be envisioned as a neural architecture with one or more hidden layers $H^{l}$. The successive layer $H^{l+1}$ can be represented as the output of an \textit{activation function} $f()$ which takes the current hidden layer $H^{l}$ and the adjacency matrix $A$.

\begin{equation}
    H^{l+1}= f(H^{l},A)
\end{equation}

$H^{0}= X$ and $H^{L} = Z$, where $X$ is the input feature matrix, $Z$ is the final feature matrix (in node classification, $Z$ is the predicted label for each node) and $L$ is the number of layers. GNN variants modify the activation function. We use the following variants of GNNs in our experiments.

\noindent\textbf{ChebNet}: The GNN model by Defferrard et al.~\cite{cheb-conv-paper} approximates the spectral convolution filter using ChebNetshev polynomials of the diagonal matrix of eigenvalues on the input graph. The applied filters are spatially localized and thus can extract features of a node's neighbours independent of the graph size.

\noindent\textbf{GCN}: The GCN model by Kifp et al.~\cite{gcn-paper} uses a localized $1^{st}$ order approximation of ChebNet with certain re-normalization tricks to enhance performance. 

\noindent\textbf{GraphSAGE}: The GraphSAGE model by Hamilton et al.~\cite{GraphSAGE} performs spatial convolution operations on the graph by sampling a fixed number of neighbours for a node and aggregating its features.  

\noindent\textbf{AGNN}: The AGNN model by Thekumparampil et al.~\shortcite{AGNN-conv-paper} applies an attention-based mechanism over a linear propagation function to aggregate neighbourhood information. Simply put it assigns varying degrees of importance to a node's neighbours, instead of treating all neighbours uniformly. 

\noindent\textbf{ARMA}: The ARMA model by  Bianchi et al.~\shortcite{ARMA-Conv-paper} utilizes auto regressive moving average (ARMA) filters instead of polynomial filters to perform convolution operations. 

\noindent\textbf{GAT}: The GAT model \cite{GAT-Conv-paper} leverage masked self-attentional layers to assign different importance to nodes of a same neighborhood, enabling a leap in model capacity.

Popular GNN models like GCN and GraphSAGE have been frequently used in the literature for node classification \cite{Graph-Conv-paper} and hence they been employed in this work. In fact, the GraphSAGE model was leveraged by Ribeiro et al.\shortcite{ribeiro2018characterizing} to classify hateful users in Twitter and has the highest reported performance amongst all the supervised frameworks. We also experiment with the recent AGNN, ARMA  and GAT model in this study since they have been shown to outperform the GCN model on the node classification task across several datasets~\cite{AGNN-conv-paper,ARMA-Conv-paper}.

\subsection{GNN + text setup}

We adopt the same setup for the different variants of GNN. The graph is the followership network $G$ for Gab and retweet network for Twitter. In order to take advantage of the text content, the input feature to the model is set to a 100 dimensional word embedding obtained by applying doc2vec on the user document\footnote{We choose doc2vec as the input feature embedding since  it outperforms other embeddings like fasttext and GloVe in the supervised setting as shown in Table~\ref{tab:hate-gab-F1}.}. The first filter (Conv1) performs convolution on the input feature vector and produces a 32 dimensional feature vector while the second filter (Conv2) further reduces it to a 2 dimensional feature vector ($Z$). A ReLu layer is added between Conv1 and Conv2 for non-linearity. We pass the final feature vector through a log-softmax layer with negative log-likelihood loss. 
 
This gives the probability of whether the user is hateful or not. The model is run for 200 epochs with Adam optimizer,  learning rate = $0.01$, weight decay = $5 \times 10^{-4}$ and dropout = 0.2. In Figure \ref{fig:text-based-classification} we have shown the schematic representation of a GNN classifier.

\section{Dataset}

For our experiments we have used the Gab~\cite{hate-speech-websci-19} and the Twitter~\cite{ribeiro2018characterizing} datasets. Gab is a social media platform which promotes itself as a \textit{``Champion of free speech''}, but has been criticized for being an echo-chamber for alt-right users~\cite{zannettou2018gab}. The site is very similar to Twitter in terms of creating posts and following others, but has loose moderation policy. Consequently, it is possible to retrieve hateful posts of users, which would have otherwise been difficult on any other platform.

Twitter, on the other hand, is a much more mainstream social media platform with stricter moderation policies. Consequently, Twitter has a significantly larger proportion of non-hateful users and closely mimics real-world distribution.

\subsection{Gab data}
\noindent\textbf{Data sampling}: We have used the Gab dataset ~\cite{hate-speech-websci-19} which comprises 381K users as well as their posts and their followership network. To ensure sufficient representation of hateful and non-hateful users, we used the sampling strategy from Mathew et al.~\shortcite{mathew2019hate}. In this sampling strategy, a lexicon of 45 high-precision hate terms are used\footnote{Lexicon available here: \url{https://goo.gl/8iHTDP}} (like `kike', `ni*ger') to identify hateful posts. An initial seed-set of 2,769 hateful users was created considering the users who have posted at least 10 such posts. Then a repost network was created  where nodes represent users and edge-weights denote the reposting frequency. This repost network is then converted to its corresponding belief network by reversing the edges and normalizing the edge-weights between 0 and 1, as outlined in \cite{hate-speech-websci-19}. Afterwards, belief score of each user was computed using a diffusion  model~\cite{golub2010naive}. An initial belief score of 1 was assigned to the hateful users and 0 to the others. Final belief values of all the users in the network were assigned after five iterations of the diffusion process. The users are then clustered on the basis of this score using $k$-means algorithm into three tiers -- \textit{``high''}, \textit{``medium''} and \textit{``low''} using a 1D KNN.
We show the distribution of the belief scores along with the three tiers in Figure~\ref{fig:belief-score-dist}.
The three tiers allow us to have better control on selecting the number of hateful and non-hateful users for the annotations, which would not be possible with random sampling.

\noindent\textbf{Annotation Guidelines}: To annotate the users as hateful we followed the hate speech definition proposed by ElSherief et al.~\shortcite{hatelingo2018}). Annotators were asked to go through all the posts of a user (rather than considering only isolated derogatory words) and use them to estimate if the account is hateful. In specific, we asked the annotators to judge if a Gab user endorses content that is \textit{humiliating, attacking or insulting, some groups or individuals based on their race, ethnicity, national origin, religion, sex, gender, sexual orientation, disability or disease.} The labelling of users was carried out by two PhD students who have extensive previous experience in hate speech annotation task. We also kept the user's information (e.g. user name, follower/followee etc.) hidden from the annotator to maintain the privacy of the users.

\noindent\textbf{Gold labels for training and evaluation}: We randomly sample 300 users from each of these three tiers with the additional constraint that the user must have posted at least 10 times. Using the annotation guidelines, the annotators manually annotate\footnote{Dubious cases which arose as a result of conflict were dropped.} each of these 900 users as hateful or non-hateful. We achieved an inter-annotator agreement of 0.772 using Cohen $\kappa$ score. On completion of the annotation and after dropping the dubious cases, the number of hateful and non-hateful users in the \textit{``low''}, \textit{``medium''} and \textit{``high''} tiers were $(57, 217)$, $(108,129)$ and $(258,29)$, respectively. This yields a final count of 423 hateful and 375 non-hateful users and constitutes our set of a total of 798 labelled instances. 

\noindent\textbf{Followership network to train GNNs}: We then construct a 1.5-degree network of these labeled users, which consists of their immediate followers, followings and connections among themselves. The nodes in the network represent the user accounts and the edges represent following relationship. A directed edge from user $A$ to user $B$ means that $A$ follows $B$. We filter the graph further by removing users with less than 10 posts. The filtered graph has 47K users and 13.8M edges and this constitutes the network $G$ required to train the GNNs for the Gab data. 

\noindent\textbf{Ethical considerations}: We only analyzed publicly available data. We followed standard ethical guidelines~\cite{rivers-ethical}, not making any attempts to track users across sites or deanonymize them. Also, taking into account user privacy, we anonymized the users information such as user name, user id etc.

\subsection{Twitter data}
We also experiment with the publicly available Twitter dataset~\cite{ribeiro2018characterizing} which followed a similar procedure to collect and label the users as hateful or non-hateful. The authors sampled and labelled 544 users as hateful and 4,427 users as non-hateful. Here the network $G$ is a retweet graph (as opposed to a followership graph) consisting of 100K nodes and 2.28M edges. A retweet graph is a directed graph $G = (V, E)$ where each node $u \in V$ represents a user in Twitter, and each edge $(u_1, u_2) \in E$ indicates that the user $u_1$ has retweeted $u_2$. Since the followership information is not present for this dataset we use the retweet network as a proxy as has also been assumed by Ribeiro et al.~\shortcite{ribeiro2018characterizing}.

\section{Experiments and results}
\subsection{Experimental setup}
To evaluate our models, we use $k$-fold stratified cross validation which can be beneficial in evaluating models having less labelled \cite{k-fold}. We set $k$ to $5$ here and for each fold, use up to 80\% dataset for training and rest 20\% for testing. Further, to simulate resource constrained setting, we take $m(\leq 80)$\% data out of the data available in each fold, for training. We report the average performance of a model across $5$-folds by varying $m$ as $5,10,15,20,50$ \& $80$. The same 20\% is always held out across all models for testing so that the comparison is fair.

\begin{table*}[ht]
 \centering
 \scriptsize
 \resizebox{\textwidth}{!}{
 \begin{tabular}{c|c|c|c|c|c|c|c||c|c|c|c|c|c}
 
 \hline
 &&\multicolumn{6}{c||}{Gab}&\multicolumn{6}{c}{Twitter}\\\hline
 Method & Inputs & 5\% & 10\% & 15\% &  20\% & 50\% & 80\% & 5\% & 10\% & 15\% &  20\% & 50\% & 80\% \\ \hline
 fastText  &$Y, X_{L}$&  0.492 &0.537 & 0.571 & 0.603 &0.690 &0.709 & 0.624  &  0.634  & 0.648  & 0.651  & 0.670  & 0.676  \\

 Glove& $Y, X_{L}$ & 0.695  & 0.720 & 0.745 & 0.750 & 0.778 & 0.784 & 0.650  &  0.666  & 0.674  & 0.681  & 0.691  & 0.695 \\

 LSTM &$Y, X_{L}$ & 0.579 & 0.600 &  0.605 & 0.608 & 0.622  &0.645 & 0.514  & 0.487  & 0.567  & 0.564 & 0.592  & 0.608 \\

 Doc2vec &$Y, X_{L}$& 0.733 &0.767 &0.783  &0.779 &0.779 &0.781 & 0.715 & 0.715 & 0.719   & 0.729 & 0.749  & 0.758 \\
  BERT &$Y, X_{L}$& 0.631 &0.660 &0.682 &0.701 &0.740 &0.764 & 0.603 & 0.665  & 0.690  & 0.709 & 0.729  & 0.740 \\
 TSVM &$Y, X$ & 0.686 & 0.704 &  0.712 & 0.712  & 0.739  & 0.753 & 0.480 & 0.520 & 0.533  & 0.533 & 0.585 & 0.611 \\
 DeepWalk &$Y,G$ & 0.652 & 0.676 & 0.700   & 0.713 & 0.723  & 0.734 & 0.757& 0.764 & 0.767 & 0.767& 0.773 & 0.779\\
Node2vec &$Y,G$ & 0.647 & 0.672 & 0.695  & 0.704 & 0.725 &0.744 & 0.692 & 0.720 & 0.732   & 0.734 & 0.749 &0.748 \\
 \hline
 GraphSAGE &$Y, X, G$ & \underline{0.778}  & \textbf{0.808} & \underline{0.806} & \underline{0.811}  & \underline{0.827} & \underline{0.828}  & \underline{0.762} & 0.773 & 0.774 & \underline{0.780}  &0.782& 0.777   \\
 GCN &$Y, X, G$ & 0.721 & 0.735 &  0.730 & 0.738  &0.751 & 0.758  & 0.756  & 0.759 &  0.767 & 0.773  &0.776 & 0.770  \\
 AGNN &$Y, X, G$ & \textbf{0.791} & \underline{0.796} &  \textbf{0.818} & \textbf{0.824} &  \textbf{0.830} & \textbf{0.833}   & \textbf{0.780} & \textbf{0.785} &  \textbf{0.785} & \textbf{0.790} &  \underline{0.786} & \textbf{0.787}   \\
 ARMA &$Y, X, G$ & 0.765& 0.778 &  0.783& 0.797& 0.809& 0.805  & 0.757  & 0.760 & 0.761  & 0.762 & 0.770  & 0.769  \\
 ChebNet &$Y, X, G$ & 0.778 & 0.802&  0.796& 0.798& 0.805& 0.812 & 0.746& 0.750&  0.754& 0.762& 0.761& 0.766\\
 GAT &$Y, X, G$ & 0.683 & 0.718&  0.725& 0.726& 0.745& 0.758 & 0.757 & \underline{0.774}&  \underline{0.781} & 0.777& \textbf{0.787} & \underline{0.782}\\ \hline
 \end{tabular}
 }
  \caption{Performance of different models for classifying users on Gab and Twitter into hateful and non-hateful  based on the mean macro F1-Score. The column x\% means that x\% of labelled instances were used for training.  $X_{L}$ represents the feature vectors of only the labelled users, $X$ represents the feature vectors of all users, $Y$ denotes the user-labels, $G$ represents the network. We perform 5-fold cross validation and note down the macro F1-Score values across the 5-folds in terms of mean. 
 The AGNN model outperforms the other models for almost all values. The best performance is marked in \textbf{bold} and the second best is \underline{underlined}.}
 	\label{tab:hate-gab-F1}
 \end{table*}
 
\subsection{Observations and insights}

\subsubsection{Observations}We illustrate the performance of the different hateful user detection models on the Gab and the Twitter data in terms of macro F1-score\footnote{We experimented with other metrics such as accuracy and observed a similar trend for both datasets. We leverage the macro F1 score to account for the high class imbalance in Twitter.} in Table \ref{tab:hate-gab-F1}.

\noindent\textit{GNNs vs text classifiers}: We observe that GNNs which combine both textual and network features exhibit an improved performance over the individual text based classifiers and the network embeddings. Amongst the GNNs, AGNN almost always boasts of the highest performance, both in terms of accuracy and macro F1-score across different amounts of labelled instances.

\noindent\textit{GNNs performance}: The attention mechanism which assigns varying importance to the nodes' neighbours accounts for the improved performance of AGNN model over other GCN model. It is to be noted that the followership/retweet network $G$ is orders of magnitudes larger\footnote{The Gab followership network and the Twitter retweet network have 13.8M and 2.28M edges as opposed to Cora (https://relational.fit.cvut.cz/dataset/CORA) and Pubmed (https://linqs.soe.ucsc.edu/data) which have only 5K and 44K edges, respectively.} than the conventional networks on which experimental results have been reported earlier and hence first order approximation may be inadequate in this setting. A similar argument holds for the improved performance of AGNN over GraphSAGE. GraphSAGE~\cite{GraphSAGE} essentially performs a linear approximation of a localized spectral convolution and is similar to the GCN model by Morris et al.~\cite{Graph-Conv-paper} barring a normalization constant.

\noindent\textit{Text classifier performance}: We observe that the Doc2vec model performs reasonably well in a supervised setting, particularly with small amount of training data. However, by including the unlabelled instances, we notice a drop in performance of TSVM model with doc2vec feature vector.

On the other hand, models which fine-tune the parameters based on the available training data, such as LSTM, BERT, DeepWalk, node2vec and GNNs show an improved performance with the increase in amount of labelled instances. 
Nevertheless, the ability of GNN models (AGNN) to achieve 0.79 macro F1-score with only 5\% labelled instances for Gab and 0.78 macro F1-score for Twitter justifies the use of GNN for detecting hateful users.

\subsubsection{Insights}
We aimed at comparing different classifiers to detect hateful users. Based on the results, it seems that GNN (especially AGNN) can leverage the network and the text features to improve the performance of this task. In order to understand why this works, we perform a detailed error analysis on the best model using text based features, i.e., doc2vec+LR and the best model using text with network features, i.e., AGNN. In particular, we focus on two situations -- \textbf{(i)} where AGNN predicts the hateful users correctly but doc2vec+LR fails (AGNN wins), \textbf{(ii)} doc2vec+LR predicts correctly but AGNN fails (doc2vec wins). For this purpose, we sampled at most 20 users for each situations from the predictions obtained in the Twitter and the GAB networks. For interpretation of the results of doc2vec+LR we use the LIME explainer~\cite{lime} to get the top 10 important words used for prediction. Similarly, for GNN, we use GNN explainer~\cite{ying2019gnnexplainer} which is used to return the top 10 most influential nodes (read users). We further annotate these users manually into hate and non hate class. We note our observations below. 

\noindent\textbf{AGNN wins}: We first consider the cases where the AGNN predicts correctly but the doc2vec does not. As a first step, from the posts of each user, we find the \% of posts having words from the hate lexicon (HL posts) \cite{mathew2019hate}. This was found to be ~2\%. In contrast, the cases where doc2vec is typically successful, the proportion of HL posts was ~5\%. Hence, it can be speculated that the representation by doc2vec cannot capture the hate dimension properly. Manually checking the top 10 words from the posts of each user, reveals that it is capturing wrong features from the users posts. For example, some such words are `fam', `girlfriend', `earthnext', `pack' etc. On the other hand, AGNN is able to make correct predictions as the user (to be classified) has several hateful neighbors in its vicinity. On average for the user being classified, the 10 influential nodes returned by the GNN explainer have 7 of them hateful. The attention on this neighborhood allows the model to learn a signature characteristic of the hateful users which makes the predictions successful.

\noindent\textbf{Doc2vec wins}: Next we consider the cases where doc2vec predicts correctly. We explore the top 10 influential nodes for these users as returned by the GNN explainer. On average 6 of these nodes (read users) are non hateful. The dominance of the non hateful users in the neighbourhood of these users actually hampers the predictions made by the AGNN models. This highlights the fact that GNN based classification will be less beneficial while detecting isolated hateful nodes (read user) in the network. Since such models do not only rely on the textual representation of an user but also on its neighbourhood, the natural expectation is that homophily would be prevalent in the network.

\noindent\textbf{Contribution of the network}: In order to understand the role of network properties in the classification task, we perform an additional experiment on the Gab data (since we have the actual follower-followee network). In this setup, we take a subset of test set with only the hate users. We use the AGNN model to predict the class on this subset. However, instead of using the doc2vec embeddings of the hate user, we average the doc2vec embeddings of the non hate users from the test set and provide it as the input user embeddings to the AGNN model. If the AGNN model is more reliant on the doc2vec embedding, then one would expect to see a lot of misclassification on the hate users test set. Please note that the network part of the hate user remains the same and only the doc2vec user embedding of the test set user is changed.

The results seem to suggest that just using the network information the AGNN is able to (re)produce 51\% of all the correct hate class predictions as would have been obtained by using the actual hate user embeddings. We perform the same experiment with the non hate users and change their emdedding to the average doc2vec embeddings of the hateful users. In this case, AGNN is able to produce the correct class only for 7\% of the non hateful users out of all that would have been produced if the actual non hateful user embeddings were used. This shows that the hateful users have a discriminative neighborhood structure and the AGNN model benefits by attending to this structure.

\begin{table}
 \centering
 \footnotesize
 \begin{tabular}{|c|c|c|c|c|c|c|}
\hline
\textbf{Methods} & \textbf{Train} & \textbf{Test} & \textbf{F1} & \textbf{F1 ($\mathcal{H}$)} & \textbf{P ($\mathcal{H}$)} & \textbf{R ($\mathcal{H}$)} \\ \hline
AGNN & \multirow{2}{*}{Twitter} & \multirow{2}{*}{Gab} & 0.75 & \textbf{0.81} & 0.71 & \textbf{0.94} \\ \cline{1-1} \cline{4-7} 
Doc2vec & & & \textbf{0.77} & 0.79 & \textbf{0.76} & 0.77 \\ \hline
AGNN & \multirow{2}{*}{Gab} & \multirow{2}{*}{Twitter} & \textbf{0.74} & \textbf{0.54} & \textbf{0.58} & \textbf{0.50} \\ \cline{1-1} \cline{4-7} 
Doc2vec &  &  & 0.58 & 0.31 & 0.22 & 0.45 \\ \hline
\end{tabular}%
\caption{Results for zero-shot cross platform evaluation. $\mathcal{H}$: Hate class, P: Precision, R: Recall.}
\label{tab:transferLearing}
\end{table}

\section{Cross platform evaluation}
Having observed the superior performance of the GNN classifiers for both the Gab and Twitter data individually, we experiment whether such models are generalizable across platforms. In particular, we train the best performing GNN classifier on one particular dataset, say GAB, and measure its performance on the other, i.e., Twitter in a zero-shot setting. We also compare these results with the best performing text based model which is doc2vec+LR. In Table \ref{tab:transferLearing}, we observe that both the GNN and doc2vec classifiers have a F1-score of around ~0.8 which tells us that the textual features learnt from user profiles in the Twitter dataset are generalizable enough. The network provides a slight benefit to the overall performance especially for the hate class. The contribution of the network is more visible when we use the Gab dataset for training. The doc2vec+LR trained on the user profiles in the Gab network performs badly when evaluated on user profiles from Twitter. In this case, the network based system seems to perform very well. This is since users in Twitter seem to use offensive words in a non-hateful sense also and hence the number of posts having words from the hate lexicon in the posts of hate and non hate users is $\sim 1$ on average. So doc2vec representation alone cannot detect the hate users. In this case, the network neighborhood structure of hateful users learnt from the Gab network seem to be quite useful for the detection.

\begin{figure}[h]
  \centering
	\includegraphics[width=\linewidth]{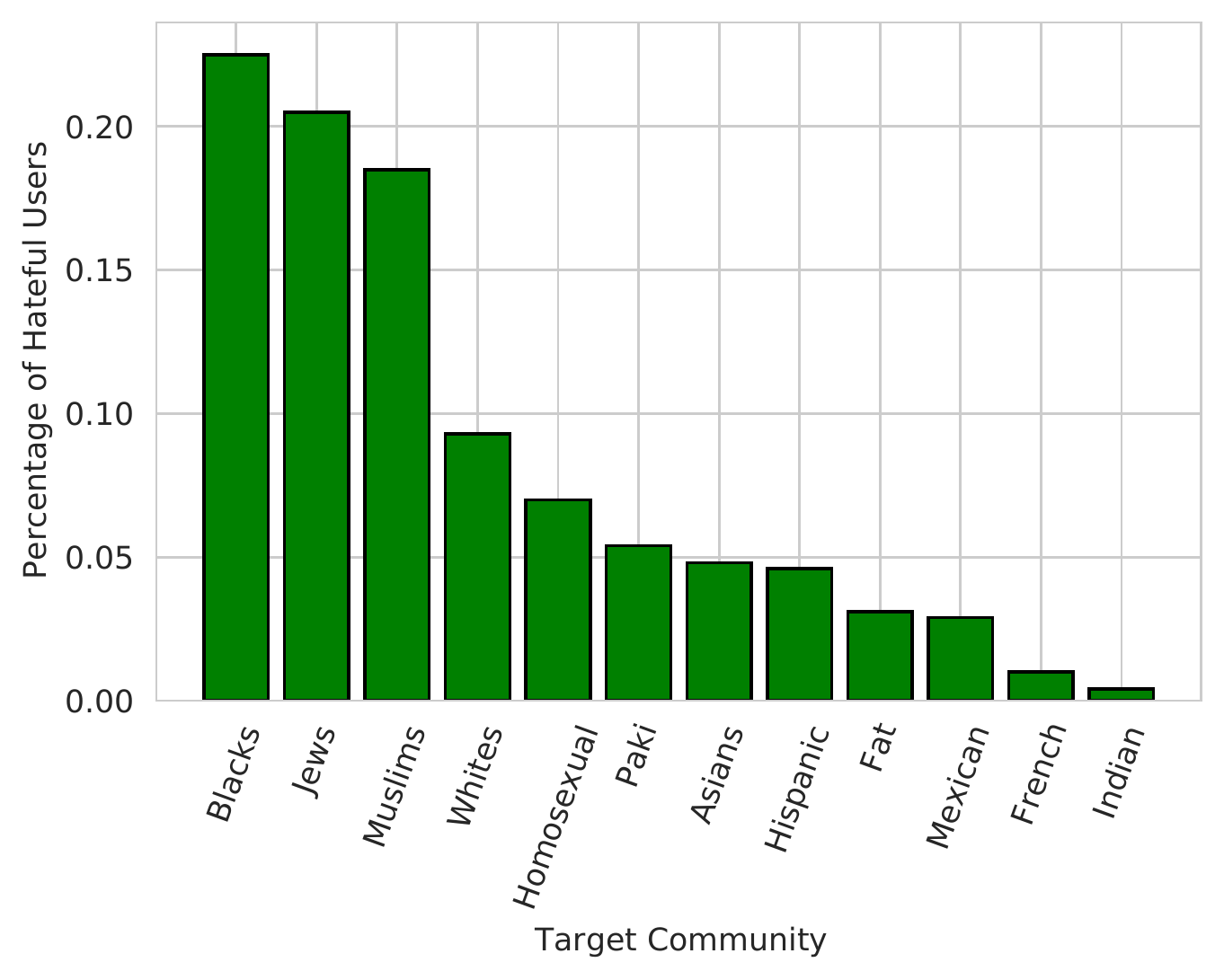}
  \captionof{figure}{Notable target communities.}
  \label{fig:hateful_target_distri1}
\end{figure}

\begin{figure}[h]
  \centering
  \includegraphics[width=\linewidth]{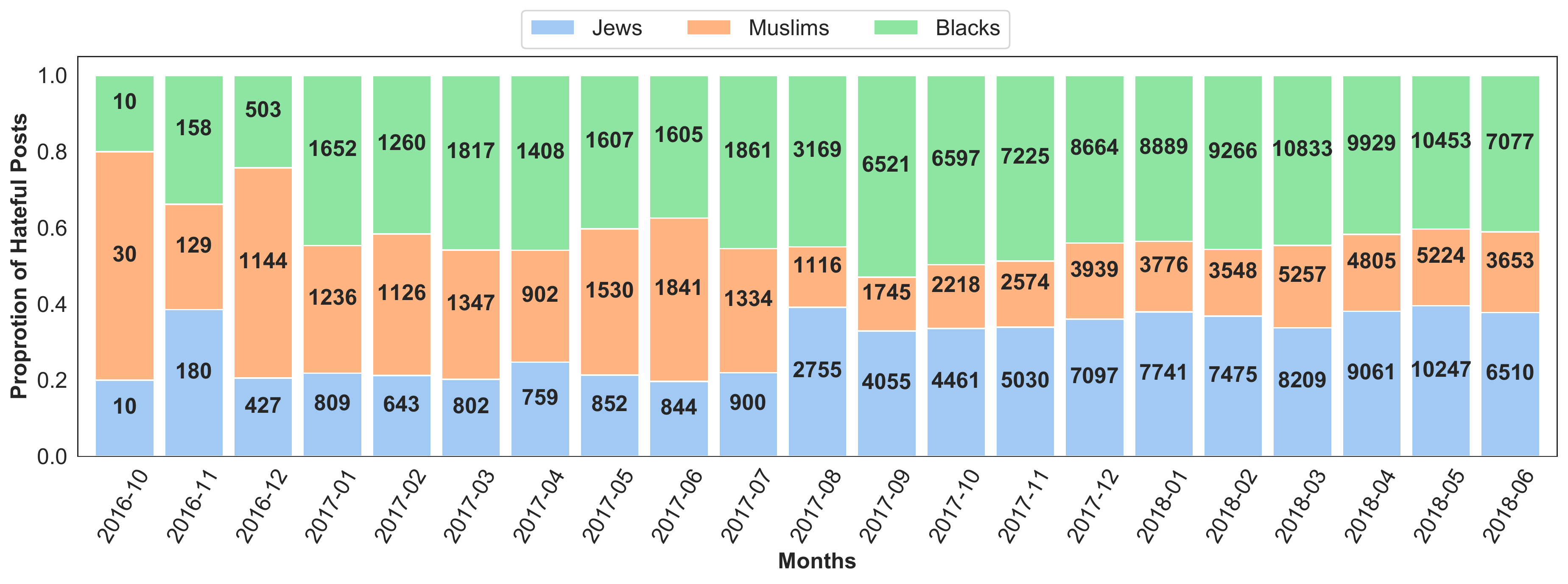}
  \captionof{figure}{Distribution of hateful posts made by the identified hateful users of a month to the three target communities -- Blacks, Muslims and Jews.}
  \label{fig:hateful_user_post}
\end{figure}

\begin{figure}[h]
  \centering
	\includegraphics[width=\linewidth]{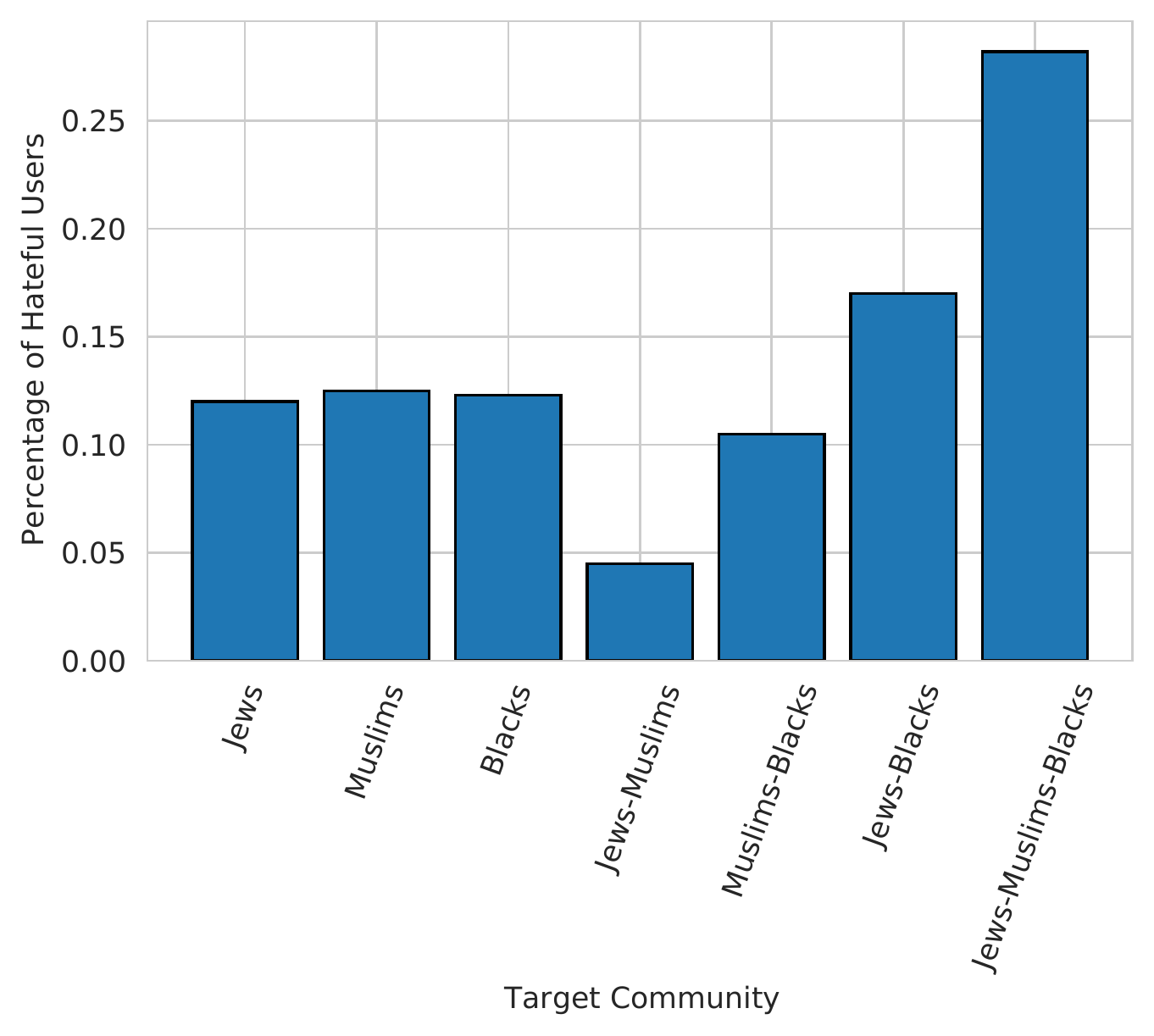}
    \caption{Distribution of the top three joint target communities.}
	\label{fig:topthree_hateful_target_distri}
\end{figure}

\begin{figure}[h]
  \centering
  \includegraphics[width=\linewidth]{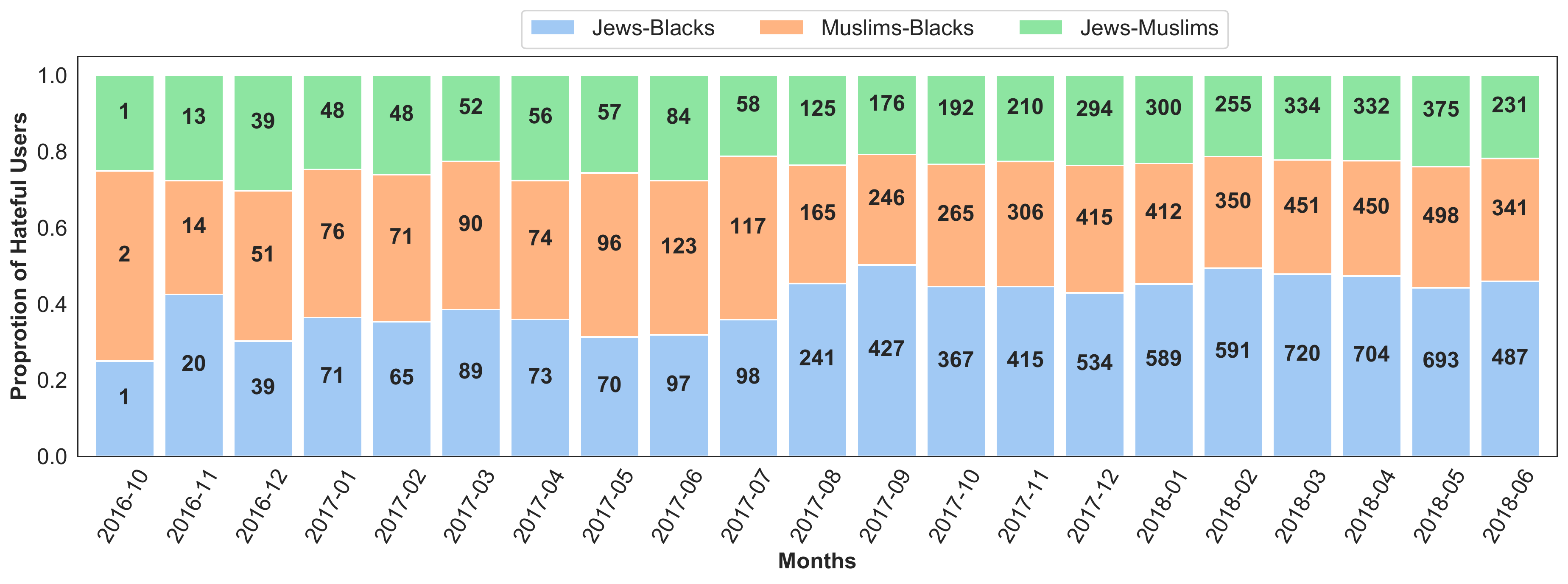}
  \captionof{figure}{Distribution of the notable joint targeted communities, i.e., `Jews-Blacks', `Muslims-Blacks' and `Jews-Muslims' per month based on the number of hateful users who targeted these joint communities.}
  \label{fig:joint_hateful_users}
\end{figure}

\section{Post facto analysis}

In  this  section, we investigate the evolution of hateful users in  Gab over time.

The precise reasons for choosing the Gab dataset for this analysis are - (i) availability of the full longitudinal data including the temporal snapshots of the followership network and (ii) loose moderation policies of the platform that enables the use of high precision keywords for obtaining reasonable results, which is not true for Twitter.

We divide our entire dataset into 21 snapshots ranging from October 2016 to June 2018 by following the snapshot generation mechanism explained in~\cite{mathew2019temporal}. They utilized a heuristic  \cite{meeder2011we} which allows to get a lower bound on the following link creation date. For each user, we note his/her followers and followings, as well as her posts on a monthly basis. We again impose the constraint that each user in the network has posted at least 10 times over her/his whole account age, to ensure that the user is sufficiently represented through the posts. Each monthly snapshot consists of users with all his/her posts and the set of followers and followees till that particular month.

We  take  the  best-performing AGNN model trained on the  entire Gab data and use it to label the users present in each snapshot as hateful or not. Once a user is labelled hateful, the user is permanently marked as hateful for the subsequent months, since the subsequent months also carry information of the current month\footnote{There might be some users who could possibly become non-hateful from hateful but this hypothesis can be safely ruled out for a platform like Gab.}. We randomly select 10 hateful users predicted by the model and manually validate them, and find 8 of them turns out to be indeed hateful. We attempt to answer the following research questions using the machine-labelled data.

 \begin{itemize}
  \item What are the target communities of these hateful users? 
  \item What is the distribution of hateful users/posts targeting toward a community?
 \end{itemize}

To answer these questions we leverage the set of high-precision lexicon obtained from Mathew et al.~\cite{mathew2019temporal} where each keyword is a derogatory slur. These keywords are then categorized\footnote{\url{https://bit.ly/3BmJqBk}} into different communities that they target. The categories are assigned through manual inspection by the authors and in consultation with the urban dictionary\footnote{\url{https://www.urbandictionary.com}} and hatebase\footnote{\url{https://hatebase.org}}. For example, ``n*gger”, ``coon”, and ``porch-monkey” are all derogatory terms used to describe blacks. We say a hateful user targets a particular community if any of his/her posts mentions any of the aforementioned keywords associated with the community. We observe that 80\% of the hateful users have used at least one of the keywords\footnote{We note that just by using the lexicon on the entire set of users might give false positives.}.

\noindent\textbf{The predominant targets}: In Figure \ref{fig:hateful_target_distri1}, we compute the distribution of targets of the hateful users using the lexicon. We observe that `Jews', `Muslims' and `Blacks' are the most prominent targets. We thus restrict our analysis to these three communities.

\noindent\textbf{The rise and rise of hatred}: In Figure \ref{fig:hateful_user_post}, we plot the distribution of the number of hateful posts against a particular target community. We observe a rise in the gross number of posts over time, highlighting that hatred is on the rise \cite{mathew2019temporal} as the website grows older. We observe that these three communities are almost equally targeted till July 2017; afterwards `Jews' and `Blacks' become slightly more prominent targets\footnote{We looked at the user distribution as well and it was similar to the post distribution.}.

\noindent\textbf{Joint targets}: To probe deeper, we observe users who target multiple communities. We state a given hateful user has targeted multiple communities if he/she uses keywords belonging to more than one community. This could be within a single post or different posts as well. We observe that a large fraction of users attack all the three communities in their posts. Figure \ref{fig:topthree_hateful_target_distri} shows the overall distribution of the hateful users targeting one or multiple communities. These categories are mutually exclusive, such that if one user targets both `Jews' and `Muslims' (`Jews-Muslims'), she is not counted in the respective sub-communities. Figure \ref{fig:joint_hateful_users} plots the temporal distribution of multi-community hatred. `Blacks-Jews’ are the most targeted communities, followed by `Muslims-Blacks' and `Jews-Muslims'.

\noindent\textbf{Trending hashtags}: Like Twitter, in Gab also \textit{hastags} can indicate the topics being discussed by the users. We attempt to find the hashtags among the posts of the hateful users which could help us to understand what kind of hate speech is being spread by the hateful users and if these are correlated to certain offline events. We collect the hashtags and their frequency for each of the 21 snapshots. Next we find the \textit{trending hashtags} for each month by finding the hashtags  which are frequent in the current month but have been infrequent in the previous month. In December 2016 some of the trending hashtags that we observe are \textit{\#BanIslam}, \textit{\#StopWhiteGenocide} which might have originated in response to the \textit{Berlin Attack}\footnote{\url{https://en.wikipedia.org/wiki/2016_Berlin_truck_attack}} that took place in December 2016 and was a part of Islamic terrorism. In January 2017, we observe a lot of incitement around the Chicago torture event done\footnote{\url{https://en.wikipedia.org/wiki/2017_Chicago_torture_incident}} that was due to a few black individuals.  Previously we observed a jump in the hate speech against `Jews' in the month of August 2017; upon investigation we find hashtags  \textit{\#UniteTheRight} and \textit{\#Charlottesville} which can be linked to the \textit{Unite the Right Rally}\footnote{\url{https://en.wikipedia.org/wiki/Unite_the_Right_rally}}. The hate users even react to the cultural events by the target community like \textit{\#BlackHistoryMonth} and call the `Blacks' as violent. Another interesting observation is the support among hate users for Tommy Robinson\footnote{\url{https://en.wikipedia.org/wiki/Tommy_Robinson_(activist)}}, an anti-Islamic activist when he was arrested in the month of June 2018 by making the hashtag \textit{\#FreeTommyRobinson} trending.

\section{Limitations and future Work}

There are a few limitations of our work. Though, AGNN model performs well by making use of both the textual and network features for classifying a user as hateful, one of the limitations of this model is it can fail for the cases where the hateful nodes have more non-hate nodes in their neighbourhood or vice-versa. Further, if a hateful user is less connected to the hateful network, network features cannot be utilized for classifying that user. Also, to find out the hateful users targeting a particular community, we have used the high-precision lexicon where each keyword is a derogatory slur. Using this method it is possible to miss one or more target communities if no derogatory slurs are used by any of the hateful users to refer to these communities.

As part of the future work, we plan to study the users who are less connected to other hateful users, and identify techniques to detect them. One can link such users to the other hate users based on the target they hate. Hence, we plan to develop a model which will not only detect hateful users but also detect the target communities of these hateful users. Another direction could be user based monitoring and possibly red alerting potential hateful users.

\section{Conclusions}
In this work, we detected hateful users on Gab and Twitter dataset using supervised and semi-supervised machine learning models. GNNs that exploit both the textual features and social connections of the users significantly outperform other models; the best model achieves macro F1-score of 0.791 on Gab and 0.780 on Twitter using only 5\% of labeled data. In order to understand the models further, we performed a detailed error analysis on doc2vec and AGNN which are the best performing models using text and text+network features, respectively. We found that doc2vec usually does not perform well when the number of hateful words in the users' post is low. In such cases the neighbourhood of a user helps the AGNN model to make correct predictions. We also notice that structural signatures learnt from a network are transferable in a zero shot setting to an unseen dataset. We perform an extensive post-facto analysis to identify how hateful posts and hateful users target different communities.

\bibliographystyle{ACM-Reference-Format}
\balance
\bibliography{sigconf}

\end{document}